\documentstyle[axodraw]{article}          
\catcode`@=11
\renewcommand{\theequation}{\thesection.\arabic{equation}}
\@addtoreset{equation}{section}
\catcode`@=12

\title{\centerline{\small SINP/TNP/99-15 \hfill hep-ph/9905206}\bigskip
\bf Faraday effect~: a field theoretical point of view}

\author{\bf Avijit K. Ganguly$^a$, 
Sushan Konar$^b$, 
Palash B. Pal$^a$\thanks{e-mail addresses: avijit@tnp.saha.ernet.in,
sushan@iucaa.ernet.in, pbpal@tnp.saha.ernet.in}\\ 
\normalsize 
$a)$ Saha Institute of Nuclear Physics, 1/AF, Bidhan-Nagar, 
Calcutta 700064, India
\\
\normalsize
$b)$ Inter-University Centre for Astronomy and Astrophysics, Pune 411007, 
India\\ 
\normalsize 
}

\date{April 1999 / Modified June 1999}

\begin{document}

\maketitle

\begin{abstract} \noindent\small 

We analyze the structure of the vacuum polarization tensor in the
presence of a background electromagnetic field in a medium. We use
various discrete symmetries and crossing symmetry to constrain the
form factors obtained for the most general case. From these symmetry
arguments, we show why the vacuum polarization tensor has to be even
in the background field when there is no background medium. Taking
then the background field to be purely magnetic, we evaluate the
vacuum polarization to linear order in it. The result shows the
phenomenon of Faraday rotation, i.e., the rotation of the plane of
polarization of a plane polarized light passing through this
background. We find that the usual expression for Faraday rotation,
which is derived for a non-degenerate plasma in the non-relativistic
approximation, undergoes substantial modification if the background is
degenerate and/or relativistic. We give explicit expressions for
Faraday rotation in completely degenerate and ultra-relativistic
media.
\end{abstract}

\section{Scope and outline of the paper}\label{sc}
It is well known that electromagnetic wave propagating through a
medium in an ambient magnetic field suffers Faraday rotation, i.e.,
the plane of a plane polarized light rotates as it travels through the
medium in the magnetic field. The amount of this rotation is derived
in various texts on electromagnetic theory \cite{emtext} and plasma
physics \cite{plasmatext} with the assumptions that the medium
consists of non-relativistic and non-degenerate electrons and
nucleons.

Faraday rotation is extensively used in a variety of situations,
including astrophysical and cosmological ones
\cite{plasmatext,interest}.  In such situations, either of the
aforesaid assumptions about the medium may not be valid. For example,
for compact stars, the plasma is likely to be degenerate. In the very
early universe, when the temperature was high, the assumption of
non-relativistic plasma is bound to break down. Motivated by such
situations, we reinvestigate this problem. For a general framework,
the formalism of quantum field theory proves to be helpful. The aim of
this paper is to use quantum field theoretical formalism to calculate
Faraday rotation in different kinds of media.

The paper is organized as follows. In Sec.~\ref{ff}, we introduce the
vacuum polarization tensor and find its most general form in a
background medium and in the presence of a general electromagnetic
field, consistent with Lorentz and gauge invariances. Calculation of
the vacuum polarization tensor requires the electron propagator, which
is discussed in Sec.~\ref{pr}. We summarize there how Schwinger's
proper-time propagator in a constant magnetic field is modified in the
presence of a background medium. In Sec.~\ref{ds}, we show how some
discrete symmetries help us constraining some of the form factors
appearing in the vacuum polarization. Following this, we set up the
calculation in Sec.~\ref{ca}. Starting from the basic Feynman rules,
we arrive at an expression for the polarization tensor that is
explicitly gauge invariant. In Sec.~\ref{dr}, we derive the expression
for Faraday rotation per unit length in terms of the components of the
polarization tensor. Then in Sec.~\ref{bk}, we provide explicit
results for different kinds of backgrounds. This is the section which
contains the essential results of the paper. The non-relativistic and
non-degenerate case is shown in Sec.~\ref{nrbk}, where we obtain the
result usually quoted in textbooks. In Secs.~\ref{dgbk} and
\ref{urbk}, we find results for a completely degenerate medium and an
ultra-relativistic one. Finally, we present our conclusions.

\section{Form factors in the polarization tensor}\label{ff}
The classical action of a free electromagnetic field is given by
\begin{eqnarray}
{\cal A} = - \, {1\over 4} \int d^4x\; F_{\lambda\rho}(x)
F^{\lambda\rho}(x) \,. 
\label{action}
\end{eqnarray}
In the momentum space, this can be written as
\begin{eqnarray}
{\cal A} = \int {d^4k \over (2\pi)^4} \; {\cal L} \,,
\end{eqnarray}
where $\cal L$ is the momentum-space Lagrangian, which can be
obtained by taking Fourier transforms in Eq.~(\ref{action}):
\begin{eqnarray}
{\cal L} = -\, {1\over 2} k^2 \widetilde g_{\lambda\rho} 
 A^\lambda (k) A^\rho(-k) \,,
\end{eqnarray}
where
\begin{eqnarray}
\widetilde g_{\lambda\rho} 
= g_{\lambda\rho} - {k_\lambda k_\rho \over k^2} \,.
\label{gtil}
\end{eqnarray}

Once quantum corrections are added, one obtains more quadratic
terms in the Lagrangian. These are represented by the vacuum
polarization tensor $\Pi_{\lambda\rho}$. In other words, after
the quantum corrections are put in, the quadratic part of the
Lagrangian becomes
\begin{eqnarray}
{\cal L} = {1\over 2} \left[ - k^2 \widetilde g_{\lambda\rho} 
+ \Pi_{\lambda\rho} (k) \right]  A^\lambda (k) A^\rho(-k)
\,.
\label{L0+Pi}
\end{eqnarray}
Owing to gauge invariance, $\Pi_{\lambda\rho}$ satisfies the conditions
\begin{eqnarray}
k^\lambda \Pi_{\lambda\rho} (k) = 0 \,, \quad 
k^\rho \Pi_{\lambda\rho} (k) = 0 \,.
\label{kpi=0}
\end{eqnarray}
In addition, Bose symmetry implies
\begin{eqnarray}
\Pi_{\lambda\rho} (k) = \Pi_{\rho\lambda} (-k) \,.
\label{bose}
\end{eqnarray}

In the vacuum, the tensor $\Pi_{\lambda\rho}$ depends only on the
momentum vector $k$. Thus, the most general form for
$\Pi_{\lambda\rho}$ is given by
\begin{eqnarray}
\Pi_{\lambda\rho} = \Pi_0 \left[ k^2 g_{\lambda\rho} - k_\lambda
k_\rho \right] \,, 
\label{vacpi}
\end{eqnarray}
where $\Pi_0$ is a Lorentz-invariant form factor, which can
therefore depend only on $k^2$. The important point is that the
tensor structure for $\Pi_{\lambda\rho}$ is exactly 
the same as the tensor
appearing in the classical Lagrangian. Thus, this correction term
can be done away with by a redefinition of the photon field
$A^\lambda$.

In a nontrivial background, this is no more the case. Although
$\Pi_{\lambda\rho}$ still has to satisfy Eq.\ (\ref{kpi=0}), the
form given in Eq.\ (\ref{vacpi}) does not follow. This is because
$\Pi_{\lambda\rho}$ can now depend, apart from the momentum
vector $k^\lambda$, on various vectors or tensors which
characterize the background medium. Even for a homogeneous
and isotropic medium, there is an extra vector in the form of the
velocity of its center of mass, $u^\lambda$. The most general
form for $\Pi_{\lambda\rho}$ in the presence of these two vectors
has been discussed in the literature \cite{NPeps}.

Our interest lies in a more complicated background where in addition
to the medium there is also an external electromagnetic field
$B_{\lambda\rho}$. We will work in the weak field limit
throughout. This means that the background field will be considered
feeble, and we will keep only linear terms in it. For the moment, we
will not specialize to magnetic fields. We will keep the discussion
general, with a medium characterized by the vector $u^\lambda$ and a
background electromagnetic field $B_{\lambda\rho}$ in arbitrary
direction.

There are many independent tensors constructed out of
$k^\lambda$, $u^\lambda$ and $B_{\lambda\rho}$ which satisfy Eq.\
(\ref{kpi=0}). For future convenience, we categorize them
into several groups. In the first group, there is only one tensor
which depends only on the vector $k^\lambda$, viz.,
\begin{eqnarray}
P^{(0)}_{\lambda\rho} \equiv \widetilde g_{\lambda\rho} 
= g_{\lambda\rho} - {k_\lambda k_\rho \over k^2} \,.
\label{P0}
\end{eqnarray}
This, of course, is the same tensor which appears in Eq.\
(\ref{vacpi}).

In the second group, we include tensors constructed only of $k$
and $u$. These are~\cite{NPeps}:
\begin{eqnarray}
P^{(1)}_{\lambda\rho} &=& \widetilde u_\lambda \widetilde
u_\rho/\widetilde u^2 \,, \label{P1}\\ 
P^{(2)}_{\lambda\rho} &=& \varepsilon_{\lambda\rho\sigma\tau}
k^\sigma u^\tau \,, 
\end{eqnarray}
where
\begin{eqnarray}
\widetilde u_\lambda = \widetilde g_{\lambda\sigma} u^\sigma \,.
\end{eqnarray}

Next, we bring in tensors constructed from $k$ and $B$ only,
without any occurence of $u$. Using the shorthand
\begin{eqnarray}
(k \cdot B)_\lambda &=& k^\sigma B_{\sigma\lambda} \,,
\end{eqnarray}
we can write these as:
\begin{eqnarray}
P'^{(1)}_{\lambda\rho} &=& k^2 B_{\lambda\rho} - k_\lambda
(k\cdot B)_\rho + k_\rho (k \cdot B)_\lambda \,, \\ 
P'^{(2)}_{\lambda\rho} &=& \varepsilon_{\lambda\rho\sigma\tau}
k^\sigma (k\cdot B)^\tau \,.
\end{eqnarray}
One might think that there might be additional terms obtained by replacing
$B_{\lambda\rho}$ by $\widetilde B_{\lambda\rho}$, where
\begin{eqnarray}
\widetilde B_{\lambda\rho} = {1\over 2}
\varepsilon_{\lambda\rho\sigma\tau} B^{\sigma\tau} \,.
\end{eqnarray}
But it is straight forward to show that no other independent term
arises this way.

Finally, to write down the tensors where all three of $k$, $u$ and $B$ 
occur, we employ a notation $(u\cdot B)_\lambda$ defined in a way 
similar to $(k\cdot B)_\lambda$. Then the tensors are:
\begin{eqnarray}
P''^{(1)}_{\lambda\rho} &=&  k\cdot u B_{\lambda\rho} - u_\lambda
	(k\cdot B)_\rho +  u_\rho (k \cdot B)_\lambda \label{P''1}\\ 
P''^{(2)}_{\lambda\rho} &=&  \varepsilon_{\lambda\rho\sigma\tau}
	k^\sigma (u\cdot B)^\tau \\ 
P''^{(3)}_{\lambda\rho} &=& \widetilde u_\lambda (k\cdot B)_\rho
	- \widetilde u_\rho (k \cdot B)_\lambda \\ 
P''^{(4)}_{\lambda\rho} &=& \widetilde u_\lambda (k\cdot B)_\rho
	+ \widetilde u_\rho (k \cdot B)_\lambda \\ 
P''^{(5)}_{\lambda\rho} &=& \widetilde u_\lambda \widetilde
	g_{\rho\tau} (u \cdot B)^\tau - \widetilde u_\rho \widetilde
	g_{\lambda\tau} (u \cdot B)^\tau \\ 
P''^{(6)}_{\lambda\rho} &=& \widetilde u_\lambda \widetilde
	g_{\rho\tau} (u \cdot B)^\tau + \widetilde u_\rho \widetilde
	g_{\lambda\tau} (u \cdot B)^\tau \\ 
P''^{(7)}_{\lambda\rho} &=& \widetilde u_\lambda (k\cdot
	\widetilde B)_\rho
	- \widetilde u_\rho (k \cdot \widetilde B)_\lambda \\ 
P''^{(8)}_{\lambda\rho} &=& \widetilde u_\lambda (k\cdot
	\widetilde B)_\rho 
	+ \widetilde u_\rho (k \cdot \widetilde B)_\lambda \\ 
P''^{(9)}_{\lambda\rho} &=& \widetilde u_\lambda \widetilde
	g_{\rho\tau} (u \cdot \widetilde B)^\tau - \widetilde
	u_\rho \widetilde g_{\lambda\tau} (u \cdot \widetilde
	B)^\tau \\  
P''^{(10)}_{\lambda\rho} &=& \widetilde u_\lambda \widetilde
	g_{\rho\tau} (u \cdot \widetilde B)^\tau + \widetilde
	u_\rho \widetilde g_{\lambda\tau} (u \cdot \widetilde
	B)^\tau \,. 
\end{eqnarray}

A collection of gauge invariant tensors which can appear in the vacuum
polarization were listed by P\'erez Rojas and Shabad
\cite{RojSha79}. They have tensors involving more than one powers of
the background field, which we are not interested in. As for the
tensors linear in $B_{\lambda\rho}$, they list what we call
$P'^{(1)}_{\lambda\rho}$, $P''^{(1)}_{\lambda\rho}$ and
$P''^{(8)}_{\lambda\rho}$, but none of the rest. We conclude that the
most general form for $\Pi_{\lambda\rho}$ consistent with Eq.\
(\ref{kpi=0}) and keeping only terms linear in the external
electromagnetic field is given by:
\begin{eqnarray}
\Pi_{\lambda\rho} = \sum_{i} \Pi^{(i)} P^{(i)}_{\lambda\rho} 
+ \sum_{i} \Pi'^{(i)} P'^{(i)}_{\lambda\rho} 
+ \sum_{i} \Pi''^{(i)} P''^{(i)}_{\lambda\rho} 
\,,
\end{eqnarray}
where in each case, the sum over $i$ runs over the appropriate
set of values. The coefficients of the tensors are form factors,
which we discuss now.

First, notice that the tensors $P^{(0)}_{\lambda\rho}$ and
$P^{(1)}_{\lambda\rho}$ do not depend on the background
electromagnetic field. The form factors associated with these terms
are related to the dielectric constant and the magnetic permeability
of the medium \cite{epsmu}. The tensor $P^{(2)}_{\lambda\rho}$ also
does not involve the background electromagnetic field. It is, however,
parity asymmetric, and accounts for natural optical activity
\cite{NPeps}. Since our aim is to discuss Faraday rotation which is
also a type of optical activity, we will disregard any natural optical
activity. Thus, we assume that the form factor associated with
$P^{(2)}_{\lambda\rho}$ is zero for our medium.

The form factors are Lorentz invariant quantities. Thus, they can
depend only on the Lorentz invariant combinations of $k^\lambda$,
$u^\lambda$ and $B_{\lambda\rho}$. Since $u^2=1$, we can obtain
the following invariant parameters, keeping at most one factor of
the background field:
\begin{eqnarray}
\omega &\equiv& k \cdot u \,,\\ 
K &\equiv& \sqrt{\omega^2 - k^2} \,,\\ 
b &\equiv& k^\lambda u^\rho B_{\lambda\rho} \,,\\ 
\widetilde b &\equiv& k^\lambda u^\rho \widetilde B_{\lambda\rho} \,.
\label{invparams}
\end{eqnarray}
In addition, of course, the form factors can depend on the
Lorentz scalars which define the background medium, e.g., the
chemical potential $\mu$ and the temperature $1/\beta$.

Since we are interested only in linear terms in the background
field, we have discarded higher order invariants in
$B_{\mu\nu}$. Moreover, notice that the tensors  
$P'^{(i)}_{\lambda\rho}$ and $P''^{(i)}_{\lambda\rho}$ are linear
in the background field. Thus, for their co-efficients, we can
neglect the field dependence for the sake of consistency. Thus,
for our purpose, the form factors $\Pi'^{(i)}$ and $\Pi''^{(i)}$
should be treated as functions of $\omega$ and $K$ only, and
possibly of $\mu$ and $\beta$. We summarize this statement as:
\begin{eqnarray}
\hat\Pi^{(i)} = \hat\Pi^{(i)} (\omega,K,\mu,\beta) \,,
\end{eqnarray}
where $\hat\Pi^{(i)}$ stands for the $\Pi'^{(i)}$'s and the
$\Pi''^{(i)}$'s. In Sec.~\ref{ds}, we will see how 
arguments about various discrete symmetries restrict the form
factors in significant ways. The constraint of hermiticity of the
action implies
\begin{eqnarray}
\Pi_{\lambda\rho} (k) = \Pi^*_{\rho\lambda} (k) \,,
\label{hermiticity}
\end{eqnarray}
whose consequences will be mentioned at the end of Sec~\ref{ds}.

	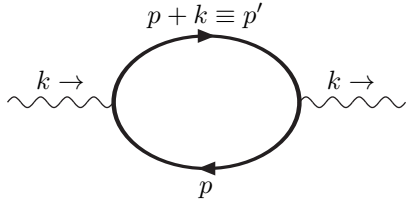
\begin{figure}
\begin{center}
\begin{picture}(150,50)(0,-25)
\Photon(0,0)(40,0){2}{4}
\Text(20,5)[b]{$k\rightarrow$}
\Photon(110,0)(150,0){2}{4}
\Text(130,5)[b]{$k\rightarrow$}
\Text(75,30)[b]{$p+k\equiv p'$}
\Text(75,-30)[t]{$p$}
\SetWidth{1.2}
\Oval(75,0)(25,35)(0)
\ArrowLine(74,25)(76,25)
\ArrowLine(76,-25)(74,-25)
\end{picture}
\end{center}
\caption[]{One-loop diagram for vacuum polarization.}\label{f:1loop}
	\end{figure}
\section{The electron propagator}\label{pr}
At the 1-loop level, the vacuum polarization tensor arises from
the diagram in Fig.\,\ref{f:1loop}. The dominant contribution to
the vacuum polarization comes from the electron line in the
loop. To evaluate this diagram, one needs to use the electron
propagator within a thermal medium in the presence of a
background electromagnetic field. Rather than working with the
complicated expression for a general background field, we will
specialize to the case of a purely magnetic field. Once this is
assumed, the field can be taken in the $z$-direction
without any further loss of generality. We will denote the
magnitude of this field by $\cal B$.

Ignoring at first the presence of the medium, the electron
propagator in such a field can be written down
following Schwinger's approach~\cite{Schwing,Tsai,Dittrich}:
\begin{eqnarray}
i S_B^V(p) = \int_0^\infty ds\; 
e^{\Phi(p,s)} G(p,s) \,,
\label{SV}
\end{eqnarray}
where $\Phi$ and $G$ are defined below. To write these in a compact
notation, we decompose the metric tensor into two parts:
\begin{eqnarray}
g_{\alpha\beta} = g^\parallel_{\alpha\beta} - g^\perp_{\alpha\beta}
\,, 
\end{eqnarray}
where
\begin{eqnarray}
g^\parallel_{\alpha\beta} &=& \mbox{diag} \big( 1,0,0,-1 \big) \,,
\nonumber \\*
g^\perp_{\alpha\beta} &=& \mbox{diag} \big( 0,1,1,0 \big) \,.
\end{eqnarray}
This allows us to write, for any two objects $q$ and $q'$ (including
the $\gamma$-matrices) carrying Lorentz indices,
\begin{eqnarray}
q\cdot q'_\parallel &=& q^{\phantom\prime}_0 q'_0 -
q^{\phantom\prime}_3 q'_3 \,, \\  
q\cdot q'_\perp &=& q^{\phantom\prime}_1 q'_1 + q^{\phantom\prime}_2
q'_2 \,. 
\end{eqnarray}
Using these notations, we can write
\begin{eqnarray}
\Phi(p,s) &\equiv& is \left( p_\parallel^2 - {\tan
(e{\cal B}s) \over e{\cal B}s} \, p_\perp^2 - m^2 \right) - \epsilon
|s| \,, 
\label{Phi}
\\
G(p,s) &\equiv&  {e^{ie{\cal B}s\sigma\!_z} \over \cos(e{\cal B}s)} \;
\left( 
\rlap/p_\parallel - \frac{e^{-ie{\cal B}s\sigma_z}} {\cos(e{\cal
B}s)}\rlap/ p_\perp + m \right) \nonumber\\*
&=& ( 1 + i\sigma_z \tan  e{\cal B}s ) 
(\rlap/p_\parallel + m ) - (\sec^2 e{\cal B}s) \rlap/ p_\perp \,,
\label{G}
\end{eqnarray}
where
\begin{eqnarray}
\sigma_z = i\gamma_1 \gamma_2 = - \gamma_0 \gamma_3 \gamma_5 \,,
\label{sigz} 
\end{eqnarray}
and we have used
\begin{eqnarray}
e^{ie{\cal B}s\sigma_z} = \cos \; e{\cal B}s + i\sigma_z \sin \;
e{\cal B}s \,.
\end{eqnarray}

The expression for $\Phi$ can have an additional gauge dependent phase
factor, but it does not contribute to the polarization
tensor. Usually, one writes $s$ instead of $|s|$ in Eq.\
(\ref{Phi}). It is equivalent since in the range of integration
indicated in Eq.\ (\ref{SV}) $s$ is never negative. However, the
definition of $\Phi(p,s)$ is useful in this form for what follows
next.

In the presence of a background medium, the above propagator is
modified to~\cite{Elmf}:
\begin{eqnarray}
iS(p) = iS_B^V(p) - \eta_F(p) \left[ iS_B^V(p) - i\overline S_B^V(p)
\right] \,,
\label{fullprop}
\end{eqnarray}
where 
\begin{eqnarray}
\overline S_B^V(p) \equiv \gamma_0 S^\dagger_V(p) \gamma_0 
\label{Sbar}
\end{eqnarray}
for a fermion propagator, and $\eta_F(p)$ contains the distribution
function for particles and antiparticles:
\begin{eqnarray}
\eta_F(p) = \Theta(p\cdot u) f_F(p,\mu,\beta) 
+ \Theta(-p\cdot u) f_F(-p,-\mu,\beta) \,.
\label{eta}
\end{eqnarray}
Here, $\Theta$ is the step function, which takes the value
$+1$ for positive values of its argument and vanishes for
negative values of the argument, and $f_F$ denotes the
Fermi-Dirac distribution function:
\begin{eqnarray}
f_F(p,\mu,\beta) = {1\over e^{\beta(p\cdot u - \mu)} + 1} \,.
\end{eqnarray}

Putting in the form of $S_B^V(p)$ from Eq.\ (\ref{SV}), we obtain the
additional term in the propagator to be
\begin{eqnarray}
S_B^\eta(p) &\equiv& 
-i \eta_F(p) \big[ S_B^V(p) - \overline S_B^V(p) \big] \nonumber\\*
&=& - \eta_F(p) \int_{-\infty}^\infty ds\; 
e^{\Phi(p,s)} G(p,s) \,,
\label{Seta}
\end{eqnarray}
with $\Phi(p,s)$ and $G(p,s)$ defined in Eqs.\ (\ref{Phi}) and
(\ref{G}).

It is straight forward to see that when ${\cal B}=0$, the
propagator in Eq.\ (\ref{SV}) reduces to 
\begin{eqnarray}
iS_0^V (p) &=& \int_0^\infty ds  \; \exp
\left[ is \left( p^2 - m^2 + i\epsilon \right) \right] \left( 
\rlap/p + m \right) \nonumber\\*
&=& i\, {\rlap/p + m \over p^2-m^2+i\epsilon} \,,
\label{S0V}
\end{eqnarray}
which is the vacuum propagator. In the same limit, the background
dependent part reduces to
\begin{eqnarray}
S_0^\eta (p) = - 2\pi \, \delta(p^2-m^2) \eta_F(p) (\rlap/p+m) \,.
\label{S0eta}
\end{eqnarray}
%

\section{Discrete symmetries and the form factors}\label{ds}
Before embarking on diagram calculations, let us discuss
some of the symmetries of the problem, which will help us
constraining some of the form factors.

\subsection{Bose symmetry}
This symmetry was already discussed in Eq.\ (\ref{bose}), viz., under
the operation
\begin{eqnarray}
k \to -k \,, \qquad \lambda \leftrightarrow \rho \,.
\label{boseop}
\end{eqnarray}
the vacuum polarization tensor must be invariant.

Since the tensors $P'^{(i)}_{\lambda\rho}$ change sign under the
operation of Eq.\ (\ref{boseop}), this implies that the
associated form factors should satisfy the condition
\begin{eqnarray}
\Pi'^{(i)} (\omega,K,\mu,\beta) = - \,\Pi'^{(i)}
(-\omega,K,\mu,\beta) \,. 
\label{Pi'bose}
\end{eqnarray}
On the form factors denoted by $\Pi''$, the effect is more
complicated since some of the tensors $P''_{\lambda\rho}$ are
symmetric under the operation of Eq.\ (\ref{bose}) and some are
antisymmetric. In general, let us write
\begin{eqnarray}
P''^{(i)}_{\rho\lambda} (-k) = n_i P''^{(i)}_{\lambda\rho} (k) 
\end{eqnarray}
where, by inspection, we see that
\begin{eqnarray}
n_i = \left\{ \begin{array}{ll} +1 \qquad & \mbox{for
$i=1,2,3,6,7,10$}\\ 
-1 & \mbox{for $i=4,5,8,9$} \,.
\end{array} \right.
\label{ni}
\end{eqnarray}
The associated form factors should then satisfy the relation
\begin{eqnarray}
\Pi''^{(i)} (\omega,K,\mu,\beta) = n_i \,\Pi''^{(i)}
(-\omega,K,\mu,\beta) \,.
\label{Pi''bose}
\end{eqnarray}

To apply this symmetry on the unprimed form factors, one has to take
into account the dependence of these form factors on $b$ and
$\widetilde b$ defined in Eq.\ (\ref{invparams}). Thus, this symmetry
implies
\begin{eqnarray}
\Pi^{(i)} (\omega,K,\mu,\beta,b,\widetilde b) &=& + \,\Pi^{(i)}
(-\omega,K,\mu,\beta,-b,-\widetilde b) \qquad \mbox{for $i=0,1$} \\ 
\Pi^{(2)} (\omega,K,\mu,\beta,b,\widetilde b) &=& - \,\Pi^{(2)}
(-\omega,K,\mu,\beta,-b,-\widetilde b) \,.
\end{eqnarray}
%

\subsection{Charge conjugation symmetry}
In calculating the form factors, we are neglecting any
corrections coming from weak interactions. In fact, these
corrections occur only at the 2-loop level, and therefore are
anyway irrelevant for the 1-loop calculation that we will be
performing. In this case, the interactions are all purely
electromagnetic, and so they obey charge conjugation (or C)
symmetry. The conclusion of this symmetry is that,
$\Pi_{\lambda\rho}$ should be invariant under the substitutions:
\begin{eqnarray}
\mu\to -\mu\,, \quad B_{\sigma\tau} \to
-B_{\sigma\tau} \,. 
\end{eqnarray}
Said in words, it means that if we calculate the vacuum polarization
in a medium with a certain background field, it should be the same as
that obtained in a charged conjugated medium with an opposite
background field. This means that, in the vacuum polarization, there
are terms even in the background field and even in $\mu$, or odd in
both. The terms linear in $B$, which we will calculate, should
therefore be odd in $\mu$. This implies that the primed form factors,
which are independent of the background medium, i.e., do not contain
$\mu$, must vanish. This is known from the direct calculations of
the polarization tensor in absence of a medium~\cite{Tsai,Urrutia}.

\subsection{A symmetry of the propagator}
Lastly, notice that in the calculation of $\Pi_{\lambda\rho}$, the
center-of-mass velocity $u^\lambda$ and the chemical potential
$\mu$ can enter only through the function $\eta_F$ in the
propagator. Further, from Eq.\ (\ref{eta}), notice that $\eta_F$
is invariant under the following transformation:
\begin{eqnarray}
u \to -\, u \,, \qquad \mu \to -\, \mu \,.
\label{accsymm}
\end{eqnarray}
So the vacuum polarization must also obey this symmetry.

For the form factors, this fact has an interesting consequence. Some
of the tensors $P''^{(i)}_{\lambda\rho}$ are even in $u$, some are
odd. Accordingly, the form factors would satisfy
\begin{eqnarray}
\Pi''^{(i)} (\omega,K,\mu,\beta) = n_i' \,\Pi''^{(i)}
(-\omega,K,-\mu,\beta) \,,
\end{eqnarray}
where
\begin{eqnarray}
n_i' = \left\{ \begin{array}{ll} -1 \qquad & \mbox{for
$i=1,2,3,4,7,8$}\\ 
+1 & \mbox{for $i=5,6,9,10$} \,.
\end{array} \right.
\label{ni'}
\end{eqnarray}

Using the consequence of C-symmetry, we can conclude that, for terms
linear in $B$,
\begin{eqnarray}
\Pi''^{(i)} (\omega,K,\mu,\beta) = - n_i' \Pi''^{(i)}
(-\omega,K,\mu,\beta) \,.
\label{C+prop}
\end{eqnarray}

In order that this is consistent with Eq.\ (\ref{Pi''bose}), we
must have
\begin{eqnarray}
n_i n'_i = -1 \qquad \mbox{(no sum on $i$).}
\end{eqnarray}
Using Eqs.\ (\ref{ni}) and (\ref{ni'}), then, we obtain that 
the doubly-primed form factors $\Pi''^{(4)}$,
$\Pi''^{(6)}$, $\Pi''^{(8)}$ and $\Pi''^{(10)}$
vanish.

The analysis performed in this section is valid for a general
electromagnetic fields in the weak limit. Although we have used
the propagator in the presence of a purely magnetic field,
substituting the more general form does not affect the
arguments. We can now discuss how this analysis can be simplified
if the background field is a purely magnetic field in the rest
frame of the background medium, i.e., if $u^\sigma
B_{\sigma\tau}=0$. Among the field-dependent tensors,
$P''^{(2)}_{\lambda\rho}$, $P''^{(5)}_{\lambda\rho}$ and
$P''^{(6)}_{\lambda\rho}$ vanish in this case. Therefore, in the
final count, we will have only four form factors associated with
the field-dependent tensors, viz., $\Pi''^{(1)}$, $\Pi''^{(3)}$,
$\Pi''^{(7)}$ and $\Pi''^{(9)}$. Since all of these are antisymmetric
tensors, the hermiticity condition of Eq.\ (\ref{hermiticity}) implies
that the corresponding form factors must be purely imaginary in the
dispersive part. 

In addition, of course, we can have the form factors $\Pi^{(0)}$ and
$\Pi^{(1)}$, whereas $\Pi^{(2)}$ vanishes because of our assumption of
vanishing natural optical activity.

\section{Calculation of the 1-loop vacuum polarization}\label{ca}
\subsection{Identifying the relevant terms}
The amplitude of the 1-loop diagram of Fig.\,\ref{f:1loop} can be
written as
\begin{eqnarray}
i \Pi_{\lambda\rho}(k) = - \int \frac{d^4p}{(2\pi)^4} (ie)^2
\; \mbox{tr}\, \left[\gamma_\lambda \, iS(p) \gamma_\rho \,
iS(p+k)\right] \,,
\end{eqnarray}
where the minus sign on the right side is for a closed fermion
loop, and $S(p)$ is the propagator given in Eq.\
(\ref{fullprop}). This implies
\begin{eqnarray}
\Pi_{\lambda\rho}(k) = -ie^2 \int \frac{d^4p}{(2\pi)^4}
\; \mbox{tr}\, \left[\gamma_\lambda \, iS(p) \gamma_\rho \,
iS(p+k)\right] \,.
\label{1loopampl}
\end{eqnarray}

From Eq.\ (\ref{fullprop}), we see that there are two terms in the
propagator --- the vacuum part $S_B^V(p)$ and the other part which
involves the background matter distribution.  If we insert two such
propagators in Eq.\ (\ref{1loopampl}), we will obtain four terms.

The term obtained from the $S_B^V$ factor in both propagators is the
contribution in the vacuum. It has no importance to our discussion of
background effects. The terms with the distribution function factor
from both propagators contributes only to the absorptive part of the
vacuum polarization, which we do not discuss in this article. Thus we
are left with the terms in which we use the vacuum part of one
propagator and the background dependent part of the other. These terms
contribute to the part $\Pi''_{\lambda\rho}$ in the notation of
Sec.~\ref{ff}. Thus
\begin{eqnarray}
\Pi''_{\lambda\rho}(k) = -ie^2 \int \frac{d^4p}{(2\pi)^4}
\; \mbox{tr}\, \Big[\gamma_\lambda S_B^\eta(p) \gamma_\rho iS_B^V(p') +
\gamma_\lambda \, iS_B^V(p) \gamma_\rho \, S_B^\eta (p') \Big] \,,
\label{SS'terms}
\end{eqnarray}
where, for the sake of notational simplicity, we have used
\begin{eqnarray}
p' = p+k \,.
\label{p'}
\end{eqnarray}
Substituting $p$ by $-p'$ in the second term and using the cyclic
property of traces, we can write Eq.\ (\ref{SS'terms}) as
\begin{eqnarray}
\Pi''_{\lambda\rho}(k) = -ie^2 \int \frac{d^4p}{(2\pi)^4}
\; \mbox{tr}\, \Big[\gamma_\lambda S_B^\eta(p) \gamma_\rho iS_B^V(p')
+ \gamma_\rho \, S_B^\eta(-p) \gamma_\lambda \,
iS_B^V (-p') \Big] \,.
\label{SS'terms2}
\end{eqnarray}
Using now the form of the propagators from Eqs.\ (\ref{SV}) and
(\ref{Seta}), we obtain
\begin{eqnarray}
\Pi''_{\lambda\rho}(k) &=& ie^2 \int \frac{d^4p}{(2\pi)^4} 
\int_{-\infty}^\infty ds \; e^{\Phi(p,s)}
\int_0^\infty ds' \; e^{\Phi(p',s')} 
\nonumber\\*
&& \times \Bigg[ \eta_F(p) \; \mbox{tr} \Big(
\gamma_\lambda G(p,s) \gamma_\rho G(p',s') \Big) + 
\eta_F(-p) \; \mbox{tr} \Big(
\gamma_\rho G(-p,s) \gamma_\lambda G(-p',s') \Big) \Bigg] \,.
\label{Pi'}
\end{eqnarray}
%

\subsection{Extracting the gauge invariant piece}
In order to discuss Faraday effect, we need only the terms in the
vacuum polarization tensor which are odd in $\cal B$. Notice that the
phase factors appearing in Eq.\ (\ref{Pi'}) are even in $\cal
B$. Thus, we need only the odd terms from the traces. Performing the
traces is straight forward, and the odd terms come out to be
\begin{eqnarray}
O_{\lambda\rho}(k) &=& 4ie^2 \int \frac{d^4p}{(2\pi)^4} \eta_-(p)
\int_{-\infty}^\infty ds \; e^{\Phi(p,s)}
\int_0^\infty ds' \; e^{\Phi(p',s')} R_{\lambda\rho} 
\end{eqnarray}
where we have introduced the notation
\begin{eqnarray}
\eta_-(p) \equiv \eta_F(p) - \eta_F(-p) \,,
\label{eta-}
\end{eqnarray}
and
\begin{eqnarray}
R_{\lambda\rho} &=& \Bigg[ \varepsilon_{\lambda\rho03} m^2 \big( \tan
e{\cal B}s - \tan e{\cal B}s' \big) 
\nonumber\\*
&& + \varepsilon_{\lambda\rho\alpha_\parallel\beta_\parallel} 
\Big( p^{\widetilde\alpha_\parallel} p'^{\beta_\parallel} \tan e{\cal
B}s - p'^{\widetilde\alpha_\parallel} p^{\beta_\parallel} \tan e{\cal
B}s' \Big)
\nonumber\\*
&& + \varepsilon_{\lambda\rho\alpha_\parallel\beta_\perp} 
\Big( p^{\widetilde\alpha_\parallel} p'^{\beta_\perp} \tan e{\cal
B}s \sec^2 e{\cal B}s' - p'^{\widetilde\alpha_\parallel}
p^{\beta_\perp} \tan e{\cal B}s' \sec^2 e{\cal B}s \Big) \Bigg] \,.
\end{eqnarray}
In writing this expression, we have used the notation of
$p^{\widetilde\alpha_\parallel}$, for example. This signifies a
component of $p$ which can take only the `parallel' indices, i.e., 0
and 3, and is moreover different from the index $\alpha$ appearing
elsewhere in the expression.

Using now the definition of $p'$ from Eq.\ (\ref{p'}), we can write 
\begin{eqnarray}
R_{\lambda\rho} = R^{(1)}_{\lambda\rho} + R^{(2)}_{\lambda\rho} \,,
\end{eqnarray}
where
\begin{eqnarray}
R^{(1)}_{\lambda\rho} = \varepsilon_{\lambda\rho\alpha_\parallel\beta}
\Big[ p^{\widetilde\alpha_\parallel} \tan e{\cal B}s + 
p'^{\widetilde\alpha_\parallel} \tan e{\cal B}s'
\Big] k^\beta 
\label{R1}
\end{eqnarray}
and
\begin{eqnarray}
R^{(2)}_{\lambda\rho} &=& \tan e{\cal B}s \Bigg[ 
m^2 \varepsilon_{\lambda\rho03} + 
\varepsilon_{\lambda\rho\alpha_\parallel\beta_\parallel}
p^{\widetilde\alpha_\parallel} p^{\beta_\parallel} +
\varepsilon_{\lambda\rho\alpha_\parallel\beta_\perp}
\Big( p^{\widetilde\alpha_\parallel} p^{\beta_\perp} +
p^{\widetilde\alpha_\parallel} p'^{\beta_\perp} \tan^2 e{\cal B}s'
\Big) \Bigg] \nonumber\\* 
&& - \tan e{\cal B}s' \Bigg[ 
m^2 \varepsilon_{\lambda\rho03} + 
\varepsilon_{\lambda\rho\alpha_\parallel\beta_\parallel}
p'^{\widetilde\alpha_\parallel} p'^{\beta_\parallel} +
\varepsilon_{\lambda\rho\alpha_\parallel\beta_\perp}
\Big( p'^{\widetilde\alpha_\parallel} p'^{\beta_\perp} +
p'^{\widetilde\alpha_\parallel} p^{\beta_\perp} \tan^2 e{\cal B}s'
\Big) \Bigg] \,.
\label{R2}
\end{eqnarray}

Obviously, $R^{(1)}_{\lambda\rho}$ is gauge invariant, i.e.,
$k^\lambda R^{(1)}_{\lambda\rho}=k^\rho R^{(1)}_{\lambda\rho}=0$. To
simplify the other term, we first note that the combinations in which
the parallel components of $p$ and $p'$ appear in Eq.\ (\ref{R2}) can
be simplified by using the following identity:
\begin{eqnarray}
\varepsilon_{\lambda\rho\alpha_\parallel\beta_\parallel}
a^{\widetilde\alpha_\parallel} b^{\beta_\parallel} = - \,
\varepsilon_{\lambda\rho03} \; a \cdot b_\parallel \,,
\end{eqnarray}
which holds for any two vectors $a$ and $b$. 
For the terms involving the transverse compoents, we make an important
observation. We will be performing the calculations in the rest frame
of the medium where $p\cdot u=p_0$. Thus, the distribution function
does not depend on the spatial components of $p$. In the last term of
each square bracket of Eq.\ (\ref{R2}), the integral over the
transverse components of $p$ has the following generic structure:
\begin{eqnarray}
\int d^2 p_\perp \; e^{\Phi(p,s)} e^{\Phi(p',s')} \times
\mbox{($p^{\beta_\perp}$ or $p'^{\beta_\perp}$)} \,.
\end{eqnarray}
Notice now that
\begin{eqnarray}
{\partial \over \partial p_{\beta_\perp}} 
\Big[ \; e^{\Phi(p,s)} e^{\Phi(p',s')} \Big] = 
{2i\over e{\cal B}} \Big( \tan e{\cal B}s \; p^{\beta_\perp} + \tan
e{\cal B}s' \; p'^{\beta_\perp} \Big)
e^{\Phi(p,s)} e^{\Phi(p',s')} \,.
\label{single_derivative}
\end{eqnarray}
However, this expression, being a total derivative, should integrate
to zero. Thus we obtain that 
\begin{eqnarray}
\tan e{\cal B}s \; p^{\beta_\perp} \stackrel\circ= - \tan
e{\cal B}s' \; p'^{\beta_\perp} \,,
\end{eqnarray}
where the sign `$\stackrel\circ=$' means that the expressions on both
sides of it, though not necessarily equal algebraically, yield the
same integral. This gives
\begin{eqnarray}
p^{\beta_\perp} &\stackrel\circ=& - \, {\tan e{\cal B}s' \over \tan
e{\cal B}s + \tan e{\cal B}s'} \; k^{\beta_\perp} \,,\nonumber\\*
p'^{\beta_\perp} &\stackrel\circ=&  {\tan e{\cal B}s \over \tan
e{\cal B}s + \tan e{\cal B}s'} \; k^{\beta_\perp} \,.
\end{eqnarray}

Using these identities, we can rewrite Eq.\ (\ref{R2}) in the
following form:
\begin{eqnarray}
R^{(2)}_{\lambda\rho} &=& 
\varepsilon_{\lambda\rho03} \Big[ 
(m^2 - p_\parallel^2) \tan e{\cal B}s -
(m^2 - p_\parallel^{\prime2}) \tan e{\cal B}s' \Big]
- \varepsilon_{\lambda\rho\alpha_\parallel\beta_\perp}
\; {\tan e{\cal B}s \; \tan e{\cal B}s' \over \tan e{\cal B}(s+s')} \;
(p+p')^{\widetilde\alpha_\parallel} k^{\beta_\perp} \nonumber\\*
&=& R^{(2a)}_{\lambda\rho} + \varepsilon_{\lambda\rho03} R^{(2b)} \,,
\end{eqnarray}
where
\begin{eqnarray}
R^{(2a)}_{\lambda\rho} &=& 
- \varepsilon_{\lambda\rho\alpha_\parallel\beta}
\; {\tan e{\cal B}s \; \tan e{\cal B}s' \over \tan e{\cal B}(s+s')} \;
(p+p')^{\widetilde\alpha_\parallel} k^\beta \,,\\
R^{(2b)} &=& (m^2 - p_\parallel^2) \tan e{\cal B}s -
(m^2 - p_\parallel^{\prime2}) \tan e{\cal B}s' 
- {\tan e{\cal B}s \; \tan e{\cal B}s' \over \tan e{\cal B}(s+s')} \;
(p+p')\cdot k_\parallel \,.
\end{eqnarray}
The term called $R^{(2b)}$ does not vanish on contraction with
arbitrary $k^\lambda$. This term is not gauge invariant, and therefore
must vanish on integration. In the Appendix, we show that this is
indeed true, so that the contribution to the vacuum polarization
tensor which is odd in $\cal B$ is given by
\begin{eqnarray}
O_{\lambda\rho}(k) &=& 4ie^2 \int \frac{d^4p}{(2\pi)^4} \eta_-(p)
\int_{-\infty}^\infty ds \; e^{\Phi(p,s)}
\int_0^\infty ds' \; e^{\Phi(p',s')} \Big[ R^{(1)}_{\lambda\rho} +
R^{(2a)}_{\lambda\rho} \Big] \nonumber\\
&=& 4ie^2 \varepsilon_{\lambda\rho\alpha_\parallel\beta} k^\beta
\int \frac{d^4p}{(2\pi)^4} \eta_-(p)
\int_{-\infty}^\infty ds \; e^{\Phi(p,s)}
\int_0^\infty ds' \; e^{\Phi(p',s')} \nonumber\\*
&\times & \Bigg[ 
p^{\widetilde\alpha_\parallel} \tan e{\cal B}s + 
p'^{\widetilde\alpha_\parallel} \tan e{\cal B}s' 
- {\tan e{\cal B}s \; \tan e{\cal B}s' \over \tan e{\cal B}(s+s')} \;
(p+p')^{\widetilde\alpha_\parallel} \Bigg] \,.
\label{Ofinal}
\end{eqnarray}
In order to perform this integral, we need to introduce further
assumptions, which will be done in Sec.~\ref{bk}.

\section{Dispersion relations}\label{dr}
\subsection{Magnetic field-independent terms in the vacuum
polarization} 
The contributions to the vacuum polarization tensor determines the
equation of motion of a photon through the medium. We have already
found the magnetic field-dependent terms in the vacuum
polarization. To obtain the dispersion relations, however, we need
also the terms which which do not depend on the background magnetic
field. These terms are necessarily even in $\cal B$ and therefore did
not appear in $O_{\lambda\rho}$. Here we outline the calculation of
these terms.

Rather than going back to Eq.\ (\ref{Pi'}) which contains also the
even terms in $\cal B$, we use directly the propagators at ${\cal B}=0$
given in Eqs.\ (\ref{S0V}) and (\ref{S0eta}) to write the background
dependent dispersive terms as
\begin{eqnarray}
\Pi'_{\lambda\rho} (k) &=& -ie^2 \int \frac{d^4p}{(2\pi)^4}
\; \mbox{tr}\, \Big[\gamma_\lambda S_0^\eta(p) \gamma_\rho iS_0^V(p') +
\gamma_\lambda \, iS_0^V(p) \gamma_\rho \, S_0^\eta (p') \Big] \,,
\label{S0S0'terms}
\end{eqnarray}
Changing, as before, the integration variable in the second term, we
obtain
\begin{eqnarray}
\Pi'_{\lambda\rho} (k) &=& -\, e^2 \int {d^4p \over (2\pi)^3}
\; {\delta(p^2-m^2) \over (p+k)^2 - m^2 } \nonumber\\*
&& \times \mbox{tr}\, 
\left[ \gamma_\lambda (\rlap/p+m) \eta_F(p) 
\gamma_\rho (\rlap/p + \rlap/k + m) +  
\gamma_\rho (\rlap/p -m) \eta_F(-p) \gamma_\lambda (\rlap/p +
\rlap/k - m) \right] \nonumber\\
&=& -4 e^2 \int {d^4p \over (2\pi)^3}
\; {\delta(p^2-m^2) \over k^2 + 2p \cdot k} 
\left[ 2p_\lambda p_\rho + p_\lambda k_\rho + k_\lambda p_\rho -
g_{\lambda\rho} p \cdot k \right] \left[ f_+ + f_- \right]\,. 
\label{00}
\end{eqnarray}
In writing the last form, we have put $p^2=m^2$ in the
denominator and in the trace, in view of the presence of the
$\delta$-function, and used
\begin{eqnarray}
\eta_F(p) + \eta_F(-p) = f_+ + f_- \,,
\end{eqnarray}
where we introduce the notations
\begin{eqnarray}
f_\pm = f_F \big( |p_0|, \mp \mu \big) \,.
\end{eqnarray}

The expression presented in Eq.\ (\ref{00}) has a particularly
simple form in the long wavelength limit, i.e., in the limit of
$K=0$. In this case, one can show that the $\Pi_{00}$ component
vanishes, whereas the $\Pi_{ij}$ components are proportional to
the unit matrix. Since the same is true for the tensor
$\widetilde g_{\mu\nu}$, we can summarize all this information by
writing 
\begin{eqnarray}
\Pi'_{\lambda\rho} (k) = \omega_0^2 \; \widetilde
g_{\lambda\rho} \,, 
\label{Pimedium}
\end{eqnarray}
where $\omega_0$ is called the plasma frequency, and is given by
\begin{eqnarray}
\omega_0^2 = 4e^2 \int {d^3p \over (2\pi)^3 2E_p} \left( 1 - {P^2
\over 3E_p^2} \right) \left[ f_+ + f_- \right] \,,
\label{ompsq}
\end{eqnarray}
where $P=|\vec p|$.

\subsection{Dispersion relations and Faraday rotation}
We have thus obtained expressions for the vacuum polarization
tensor. For the rest of this paper, we will consider only photon
propagation along the direction of the magnetic field. Thus, in Eq.\
(\ref{Ofinal}), the index $\beta$ can only take the values 0 or
3. Since the index $\alpha$ appearing in that equation had also 
parallel components only, the antisymmetric tensor now implies that 
$\Pi''_{\lambda\rho}$ vanishes unless both $\lambda$ and $\rho$ are
transverse, i.e., have values 1 or 2. Thus, the only non-vanishing
components of $\Pi''_{\lambda\rho}$ are:
\begin{eqnarray}
\Pi''_{12} (k) = - \Pi''_{21} (k) = - \, ia \,,
\label{12and21}
\end{eqnarray}
where $a$ has to be determined by evaluating the integral in Eq.\
(\ref{Ofinal}).  The contributions which come from the medium even
without the magnetic field has been given in Eq.~(\ref{Pimedium}).

To obtain the dispersion relations, we go back to the Lagrangian
given in Eq.\ (\ref{L0+Pi}). The equation of motion obtained from
this Lagrangian is 
\begin{equation}
\left[ (-k^2+\omega_0^2) \widetilde g_{\lambda\rho} +
\Pi''_{\lambda\rho} \right] A^\rho = 0 \,.
\end{equation}
In view of the Lorentz gauge condition $k_\rho A^\rho=0$, this
can also be written as
\begin{equation}
\left[ (-k^2+\omega_0^2) g_{\lambda\rho} +
\Pi''_{\lambda\rho} \right] A^\rho = 0 \,.
\label{eqmotion}
\end{equation}

For the transverse components of the photon field $A^\rho$, the
above equation implies the following condition:
\begin{eqnarray}
\left(\begin{array}{cc} 
-k^2 + \omega_0^2 & -ia  \\
ia & -k^2 + \omega_0^2
\end{array}\right)
\left(\begin{array}{c}A_1\\ A_2 \end{array}
\right) = 0 \,.
\label{d2}
\end{eqnarray}
The eigenvalues of the matrix give the dispersion relations
\begin{eqnarray}
k^2 = \omega_0^2 \pm a
\label{disp}
\end{eqnarray}
for the normalized eigenmodes
\begin{eqnarray}
\left( A_1 \pm iA_2 \right) / \sqrt 2 \,,
\end{eqnarray}
which describe circularly polarized states of the photon.

Writing $k^2$ as $\omega^2-K^2$, we obtain the following solutions for
$K$:
\begin{eqnarray}
K_\pm = \sqrt{\omega^2 - \omega_0^2} \left[ 1 \mp {a \over 
\omega^2 - \omega_0^2} \right]^{1/2}
\end{eqnarray}
For small magnetic fields, $a$ will be small, and then we can write
\begin{eqnarray}
K_\pm = \sqrt{\omega^2 - \omega_0^2} \left[ 1 \mp {a \over 2 
(\omega^2 - \omega_0^2)} \right] \,,
\end{eqnarray}
which gives, for the difference of the two solutions,
\begin{eqnarray}
\Delta K = {a \over \sqrt{\omega^2 - \omega_0^2}} \,.
\end{eqnarray}

For a plane polarized electromagnetic wave propagating with a frequency
$\omega$, this means that, after travelling a distance $\ell$,
the plane of propagation will be rotated by an amount
$\ell\,\Delta K$. Thus, the rate of rotation of the polarization
angle $\Phi$ is given by
\begin{eqnarray}
{d\Phi \over d\ell} = \Delta K = {a \over 
\sqrt{\omega^2 - \omega_0^2}} \,.
\label{dphidl}
\end{eqnarray}
This is the Faraday rotation per unit length. The magnitude of this
quantity is thus determined once we determine $a$ and $\omega_0$.%
\footnote{Instead of $\omega_0$, one can also use the index of
refraction $r$, defined by the relation $r=K/\omega$.  In absence of
the magnetic field, i.e., when $a=0$, Eq.\ (\ref{disp}) gives
\begin{eqnarray*}
r^2 = 1 - {\omega_0^2 \over \omega^2} \,.
\end{eqnarray*}
We can use this relation to eliminate $\omega_0$ from the
formulas above and express everything in terms of $r$, i.e., the
refractive index in absence of the magnetic field. 
}

In what follows, we find out what $\omega_0$ and $a$ are for different
types of backgrounds, and consequently what is the amount of Faraday
rotation sufferred by plane polarized light in such backgrounds.  We
will do this for three different kinds of backgrounds, depending on
the relative importance of the temperature $T=1/\beta$, the chemical
potential $\mu$, and the electron mass $m_e$.

\section{Results for different backgrounds}\label{bk}
\subsection{General observations and assumptions}
Before starting with any of the specific cases, let us note some
general features and some common assumptions in the
calculations. We will perform all the calculations assuming that
the background medium is at rest in the frame in which we have a
purely magnetic field. In other words, for the 4-vector $u$, the
only non-zero component is the time component, which has the
value unity. All other components are zero.

As already mentioned, we will consider photon propagation along the
$z$-direction (positive or negative). In addition, we will take the
long wavelength limit, i.e., $K\ll\omega$. This implies that in Eq.\
(\ref{Ofinal}), the term with the external factor of $k^0=\omega$
dominates over the one with $k^3=K$.

Finally, we will assume the magnetic field to be small, so that we can
use only the linear terms in $\cal B$. To this order, then, the
dominant contribution of Eq.\ (\ref{Ofinal}) is given by
\begin{eqnarray}
\Pi''_{\lambda\rho} (k) &=& 8i e^3 {\cal B} 
\varepsilon_{\lambda\rho30} \omega I \,,
\label{pi''final}
\end{eqnarray}
where
\begin{eqnarray}
I &=& 
\int \frac{d^4p}{(2\pi)^4} \eta_-(p) p_0
\int_{-\infty}^\infty ds \; e^{is(p^2-m^2)-\epsilon |s|}
\int_0^\infty ds' \; e^{is'(p^{\prime2}-m^2)-\epsilon |s'|} 
\Big[ s + s' - {ss' \over s+s'} \Big] \,.
\label{I0}
\end{eqnarray}
Here, since the other factor is already linear in $\cal B$, we have
used ${\cal B}=0$ in the exponents. Moreover, we have made the further
assumption that $\omega\ll m_e$, which enables us to neglect $k_0$
compared to $p_0$ in the factor inside the square bracket. In the
notation introduced in Eq.\ (\ref{12and21}), we can write
\begin{eqnarray}
a &=& 8 e^3 {\cal B} \omega I \,.
\label{a}
\end{eqnarray}

The expression for $I$ can be put in a convenenient form. For this,
we first define the integral
\begin{eqnarray}
J_n &=& 
\int \frac{d^4p}{(2\pi)^4} \eta_-(p) p_0
\int_{-\infty}^\infty ds \; e^{is(p^2-m^2)-\epsilon |s|}
\int_0^\infty ds' \; e^{is'(p^{\prime2}-m^2)-\epsilon |s'|} \; {s'}^n
\,. 
\end{eqnarray}
If we now rewrite the factor in the square brackets in Eq.\ (\ref{I0})
in the following form:
\begin{eqnarray}
s + s' - {ss' \over s+s'} = (s+s') - s' + {{s'}^2 \over s+s'} \,,
\end{eqnarray}
it is easily seen that
\begin{eqnarray}
I = i {\partial \over \partial (m^2)} J_0 - J_1 + i \int d(m^2)\;
J_2 \,.
\end{eqnarray}
The task is now to evaluate $J_n$ for $n=0,1,2$. The $s$-integral in
$J_n$ gives
\begin{eqnarray}
\int_{-\infty}^\infty ds \; e^{is(p^2-m^2)-\epsilon |s|} = 2\pi \,
\delta(p^2-m^2) \,,
\end{eqnarray}
whereas the $s'$-integral gives
\begin{eqnarray}
\int_0^\infty ds' \; e^{is'(p^{\prime2}-m^2)-\epsilon |s'|} \; {s'}^n
= {i^{n+1} n! \over \big( p^{\prime2} - m^2 \big)^{n+1}} \,.
\end{eqnarray}

Writing now
\begin{eqnarray}
\delta\left( p^2-m^2 \right) = {1\over 2E_p} \left[
\delta(p_0-E_p) + \delta(p_0+E_p) \right] \,,
\end{eqnarray}
and using
\begin{eqnarray}
\eta_-(p) = {\rm sign}(p_0) \big( f_+(p) - f_-(p) \big)
\end{eqnarray}
which follows from the definitions in Eqs.\ (\ref{eta}) and
(\ref{eta-}), we obtain
\begin{eqnarray}
J_n = i^{n+1} n!
\int {d^4p \over (2\pi)^3} 
{p_0 \mbox{sign}\,(p_0) \over 2E_p} \left[
\delta(p_0-E_p) + \delta(p_0+E_p) \right] 
{f_+ - f_- \over 
\left( k^2 + 2 p_0\omega - 2PK \cos \theta' \right)^{n+1}} \,.
\end{eqnarray}
Here, $k^\mu \equiv (\omega, \vec K)$, $P\equiv |\vec p|$ and
$\theta'$ is the angle between $\vec K$ and $\vec p$. We have denoted
this angle by $\theta'$ in order to emphasize that, for a general
direction of propagation, it can be different from the angle $\theta$
which is measured from the $z$-axis, i.e., from the direction of the
magnetic field which we have already specified. For our specific case
of photon propagation along the magnetic field direction, however, we
will put $\theta'=\theta$. 

We further notice that we can neglect the term $k^2$
because of our assumptions stated earlier. Thus,
\begin{eqnarray}
J_n &=& {i^{n+1} n! \over 8}
\int {d^3p \over (2\pi)^3} 
\left( f_+ - f_- \right) \left[ {1 \over 
\left( E_p\omega - PK \cos \theta \right)^{n+1}} 
+ {1 \over 
\left( -E_p\omega - PK \cos \theta \right)^{n+1}} 
\right]\,.
\label{Jn}
\end{eqnarray}
The azimuthal integration gives a factor $2\pi$, and the $\theta$
integration can be exactly performed here. This shows that $J_n=0$ for
even values of $n$. This conclusion can be avoided only if $\omega$
and/or $K$ becomes comparable to $m_e$. Since we have already assumed
otherwise, we get
\begin{eqnarray}
I=-J_1 \,,
\end{eqnarray}
i.e., 
\begin{eqnarray}
I &=& {1 \over 8}
\int {d^3p \over (2\pi)^3} 
\left( f_+ - f_- \right) \left[ {1 \over 
\left( E_p\omega - PK \cos \theta \right)^2} 
+ {1 \over 
\left( -E_p\omega - PK \cos \theta \right)^2} 
\right]\,.
\label{J1}
\end{eqnarray}

In general, however, even this integral cannot
be performed analytically. So, in order to discuss the amount of
Faraday rotation caused by this term, we need to take recourse to
some specific limits.

\subsection{Connection with the form factors}
It is of interest to see how our final result for
$\Pi''_{\lambda\rho}$ conforms to the general form obtained on
the basis of gauge and Lorentz invariance, a subject that was
discussed in Sec.~\ref{ff}. At the end of Sec.~\ref{ds}, we
remarked that in our case, we can get at most four independent
form factors, viz, those associated with the field-dependent
tensors $P''^{(1)}_{\lambda\rho}$, $P''^{(3)}_{\lambda\rho}$,
$P''^{(7)}_{\lambda\rho}$ and $P''^{(9)}_{\lambda\rho}$. However,
the simplifying assumptions made above imply that all components
of $\widetilde u_\lambda$ vanishes to the leading order. Thus,
only $P''^{(1)}_{\lambda\rho}$ survives in this case. Moreover,
since we choose the direction of propagation to be along the
magnetic field, $(k\cdot B)_{\lambda}=0$ as well. Thus, from Eq.\
(\ref{P''1}), we find that the tensor $P''^{(1)}_{\lambda\rho}$,
in the case of our choice, is simply proportional to
$B_{\lambda\rho}$. This is what the explicit calculation of Eq.\
(\ref{pi''final}) tells us as well.

\subsection{A non-relativistic background}\label{nrbk}
Suppose we have a gas of electrons and positrons where all the
particles are non-relativistic. 
In this case, we can put $E_p\approx m_e$ within the integral,
and neglect all occurrences of $P$ since it is small compared to
$E_p$. Then we obtain
\begin{eqnarray}
I &=&  {1\over 4m_e^2 \omega^2} 
\int {d^3p \over (2\pi)^3} 
\left( f_+ - f_- \right) \nonumber\\* 
&=& {1\over 8m_e^2 \omega^2} 
\left( n_e - n_{\bar e} \right) \,.
\end{eqnarray}
Using Eqs.\ (\ref{a}) and (\ref{dphidl}) now, the Faraday rotation per
unit length is obtained to be
\begin{eqnarray}
{d\Phi \over d\ell}  = 
{e^3 {\cal B} \over m_e^2 \omega \sqrt{\omega^2 - \omega_0^2}}
\; \left( n_e - n_{\bar e} \right) \,,
\end{eqnarray}
where $\omega_0$, in this limit, can be simplified by using the
general formula in Eq.\ (\ref{ompsq}):
\begin{eqnarray}
\omega_0^2 = {2e^2 \over m_e} \int {d^3p \over (2\pi)^3} 
\left[ f_+ + f_- \right] = {e^2 \over m_e} 
\left( n_e + n_{\bar e} \right) \,.
\label{ompsqnr}
\end{eqnarray}

If the background contains no positrons, the expression for
Faraday rotation can be written as
\begin{eqnarray}
{d\Phi \over d\ell}  = 
{\omega_0^2 \omega_c \over \omega
\sqrt{\omega^2 - \omega_0^2}} \,,
\label{FCnr}
\end{eqnarray}
where $\omega_c\equiv e{\cal B}/m_e$ is called the cyclotron
frequency. 

\subsection{A degenerate background}\label{dgbk}
We now consider a degenerate electron background at zero
temperature. The distribution functions are now given by
\begin{eqnarray}
f_+ &=& \left\{ \begin{array}{ll} 1 \qquad & \mbox{for $P\leq P_F$}
\\ 0 & \mbox{for $P > P_F$}
\end{array} \right. \nonumber\\*
f_- &=& 0 \,,
\label{degf}
\end{eqnarray}
where $P_F$ is called the Fermi momentum. As we know, although
the temperature is zero, the electrons need not be
non-relativistic in this case, since Pauli exclusion principle
would require all electrons to be in different states, and so
some of them can be at very large momentum. The number density of
electrons in this case is given by
\begin{eqnarray}
n_e = 2 \int {d^3p \over (2\pi)^3} = {P_F^3 \over 3\pi^2} \,.
\end{eqnarray}

In this case, we first calculate the plasma frequency. Performing
the angular integrations of Eq.\ (\ref{ompsq}), this can be
written in the form 
\begin{eqnarray}
\omega_0^2 = {e^2 m_e^2 \over \pi^2} \int_0^{x_F} dx\; \left(
{x^2 \over (1+x^2)^{1/2}} - {x^4 \over 3(1+x^2)^{3/2}}
\right) \,,
\end{eqnarray}
where $x$ is the integration variable defined by $P/m_e$, and
$x_F=P_F/m_e$. The 
integration can be performed in a straight forward manner by
substituting $x=\sinh\zeta$, and the result is
\begin{eqnarray}
\omega_0^2 = {e^2 m_e^2 \over 3\pi^2} \; {x_F^3 \over
\sqrt{1+x_F^2}} = {e^2 \over 3\pi^2} \; {P_F^3 \over E_F}
\end{eqnarray}
where $E_F$ is the Fermi energy,
\begin{eqnarray}
E_F = \sqrt{P_F^2+ m_e^2} \,.
\end{eqnarray}

We now evaluate the integral $I$. Starting from the expression in
Eq.\ (\ref{J1}) for the general case, we perform the angular
integrations to obtain
\begin{eqnarray}
I &=& {1\over 16\pi^2 K} 
\int_0^\infty dP \, P (f_+ - f_-) 
\Bigg[ {1 \over E_p\omega - PK} 
- {1 \over E_p\omega + PK} \Bigg] \nonumber\\*
&=& {1\over 8\pi^2 \omega^2} 
\int_0^\infty dP \, {P^2 \over E_p^2} (f_+ - f_-) 
\,.
\label{angularintdone}
\end{eqnarray}

Using now the distribution functions appropriate for this case from
Eq.\ (\ref{degf}), we obtain
\begin{eqnarray}
I &=& {1\over 8\pi^2 \omega^2} 
\Big[ P_F - m_e \tan^{-1} (P_F/m_e) \Big]\,,
\end{eqnarray}
where the result of the arctan function is restricted within the
domain 0 to $\pi/2$.  {}From this, we obtain the Faraday rotation per
unit length to be
\begin{eqnarray}
{d\Phi \over d\ell}  = 
{\omega_0^2 \omega_c \over \omega
\sqrt{\omega^2 - \omega_0^2}}  \cdot {3m_e E_F \over P_F^3} 
\Big[ P_F - m_e \tan^{-1} (P_F/m_e) \Big]\,.
\label{FCdg}
\end{eqnarray}

It can be easily checked that if $P_F\ll m_e$, in which case the
background is non-relativistic, the formulas derived for this
case reduce to those derived in Sec.~\ref{nrbk}.

\subsection{An ultra-relativistic background}\label{urbk}
Let us now discuss the case where the temperature $T$ of the
background is much higher than the electron mass. In this case,
we can put $E_p\approx P$. Then, using the dimensionless
integration variable $y=P/T$, the plasma frequency can be
expressed as
\begin{eqnarray}
\omega_0^2 = {2e^2 \over 3\pi^2\beta^2} \int_0^\infty dy\; y \left(
{1\over \exp (y-\beta\mu) + 1} + {1\over \exp (y+\beta\mu) + 1} 
\right) \,.
\label{F1}
\end{eqnarray}
This integral can in fact be performed exactly. In the first
integral, use the new integration variable $y'=y-\beta\mu$. In
the second one, use $y'=y+\beta\mu$. The resulting integrations
can then be written in the form
\begin{eqnarray}
\omega_0^2 = {2e^2 \over 3\pi^2\beta^2} \left[ 
\int_0^\infty dy'\; {2y'\over e^{y'} + 1} + 
\int_{-\beta\mu}^0 dy'\; {y'+\beta\mu\over e^{y'} + 1} 
- \int_0^{\beta\mu} dy'\; {y'-\beta\mu\over e^{y'} + 1} 
 \right] \,.
\end{eqnarray}
The first integral can now be performed by expressing the
denominator as a geometric series. The other two can be combined
after substituting $y'\to-y'$ in the second integral, and the
final result is
\begin{eqnarray}
\omega_0^2 
&=& e^2 \left[ {1 \over 9\beta^2} + {\mu^2 \over 3\pi^2} \right]\,.
\end{eqnarray}

For the integral $I$, we start from the general expression in Eq.\
(\ref{angularintdone}). Using similar substitutions as before, we
obtain 
\begin{eqnarray}
I &=& {1 \over 8\pi^2\beta\omega^2} \int_0^\infty dy\;  \left(
{1\over \exp (y-\beta\mu) + 1} - {1\over \exp (y+\beta\mu) + 1} 
\right) \nonumber\\*
&=& {\mu \over 8\pi^2\omega^2} \,. 
\label{F2}
\end{eqnarray}
The Faraday rotation is obtained from Eq.~(\ref{dphidl}) and
(\ref{a}):
\begin{eqnarray}
{d\Phi \over d\ell} = {e^3 {\cal B} \mu \over \pi^2\omega
\sqrt{\omega^2 - \omega_0^2}} \,.
\label{FCur}
\end{eqnarray}
If we put
$\beta\to\infty$, the results obtained for this case reduce to
those obtained in \S\,\ref{dgbk} with $m_e=0$.

As for the previous cases, one may want to express these results
in terms of the number densities of electrons and positrons in
the medium rather than in terms of the chemical potential
$\mu$. The connection comes from the relation
\begin{eqnarray}
n_e - n_{\bar e} = 2 \int {d^3p \over (2\pi)^3} \; \left( f_+ -
f_- \right) \,.
\end{eqnarray}
Again, the integration can be performed exactly, following the
steps described above, and the result is
\begin{eqnarray}
n_e - n_{\bar e} = {\mu \over 3\beta^2} + {\mu^3 \over 3\pi^2} \,.
\end{eqnarray}
One can use this to express $\mu$ in terms of $n_e - n_{\bar e}$.

\section{Conclusions}
We have thus shown that the amount of Faraday rotation depends very
significantly on the characteristics of the medium in which the
background magnetic field rests. If the medium consists of
non-relativistic particles only, we obtain the formula given in Eq.\
(\ref{FCnr}). Usually, we assume that this formula should be
applicable for low-temperature media, since the particles should be
non-relativistic in this case. However, we show that if the medium is
strongly degenerate, the formula changes. For $P_F\ll m_e$ it still
agrees with the non-relativistic result as it should. But for $P_F\gg
m_e$ the change is drastic, and Faraday rotation becomes very small,
as can be seen from Eq.\ (\ref{FCdg}). Similarly, if the medium is so
hot that the kinetic energies of the particles are much larger than
their masses, we obtain a different result, as shown in Eq.\
(\ref{FCur}). But it is interesting to note that in all the cases
discussed, the quantity $a$ has the same dependence on $\omega$, viz.,
that it is inversely proportional to $\omega\sqrt{\omega^2-\omega_0^2}$.

\bigskip
\setcounter{section}{0}
\renewcommand{\thesection}{Appendix:}
\renewcommand{\theequation}{\Alph{section}.\arabic{equation}}
\section{Proof of gauge invariance}
In the text, we claimed that the contribution coming from $R^{(2b)}$
must vanish in order that the vacuum polarization tensor is gauge
invariant. Here, we justify this claim. This contribution is
proportional to the following integral:
\begin{eqnarray}
C &=& \int \frac{d^4p}{(2\pi)^4} \eta_-(p)
\int_{-\infty}^\infty ds \; e^{\Phi(p,s)}
\int_0^\infty ds' \; e^{\Phi(p',s')} R^{(2b)} \nonumber\\
&=& \int \frac{d^4p}{(2\pi)^4} \eta_-(p)
\int_{-\infty}^\infty ds \; e^{\Phi(p,s)}
\int_0^\infty ds' \; e^{\Phi(p',s')} \nonumber\\*
&\times & \Bigg[ (m^2 - p_\parallel^2) \tan e{\cal B}s -
(m^2 - p_\parallel^{\prime2}) \tan e{\cal B}s' 
- {\tan e{\cal B}s \; \tan e{\cal B}s' \over \tan e{\cal B}(s+s')} \;
(p+p')\cdot k_\parallel \, \Bigg] \,. 
\label{Zero}
\end{eqnarray}
Using the definition of the exponential factor $\Phi(p,s)$ from Eq.\
(\ref{Phi}), we notice that
\begin{eqnarray}
m^2 \tan e{\cal B}s \; e^{\Phi(p,s)} \, e^{\Phi(p',s')} &=& 
\tan e{\cal B}s \, \Big\{i {d \over ds'} + (p_\parallel^{\prime2} - \sec^2
e{\cal B}s' \; p_\perp^{\prime2}) \Big\}  \, e^{\Phi(p,s)} \, e^{\Phi(p',s')}
\,, \\ 
m^2 \tan e{\cal B}s' \; e^{\Phi(p,s)} \, e^{\Phi(p',s')} &=& 
\tan e{\cal B}s' \, \Big\{i {d \over ds} + (p_\parallel^2 - \sec^2 e{\cal
B}s \; p_\perp^2) \Big\} \, e^{\Phi(p,s)} \, e^{\Phi(p',s')}  \,. 
\end{eqnarray}
This implies that we can write
\begin{equation}
C = C_1 + iC_2 \,,
\end{equation}
where
\begin{eqnarray}
C_1 &=& \int \frac{d^4p}{(2\pi)^4} \eta_-(p) 
\int_{-\infty}^\infty ds \; e^{\Phi(p,s)} \int_0^\infty ds' \;
e^{\Phi(p',s')} \nonumber\\* 
&\times & \Bigg[ (p_\parallel^{\prime2} - \sec^2 e{\cal B}s' \;
p_\perp^{\prime2} - p_\parallel^2) \tan e{\cal B}s - 
(p_\parallel^2 - \sec^2 e{\cal B}s \; p_\perp^2 - p_\parallel^{\prime2})
\tan e{\cal B}s' \nonumber \\* 
&& - \; {\tan e{\cal B}s \; \tan e{\cal B}s' \over \tan e{\cal
B}(s+s')} \; (p+p')\cdot k_\parallel \, \Bigg] \,, \\* 
C_2 &=& \int \frac{d^4p}{(2\pi)^4} \eta_-(p) 
\int_{-\infty}^\infty ds \; \int_0^\infty ds' \left(\tan e{\cal B}s \,
{d \over ds'} - \tan e{\cal B}s' \, {d \over ds} \right) 
\; e^{\Phi(p,s)} \; e^{\Phi(p',s')} \,.
\end{eqnarray}

Let us first consider the contribution $C_2$. Performing the
$s'$-integration in the first term and the $s$-integration in the
second, it can be written as
\begin{eqnarray}
C_2 = \int \frac{d^4p}{(2\pi)^4} \eta_-(p) 
\Bigg[ e^{\Phi(p',s')} \Bigg|_0^{\infty} 
\int_{-\infty}^\infty ds \; \tan e{\cal B}s \; e^{\Phi(p,s)} \;
- e^{\Phi(p,s)} \Bigg|_{-\infty}^{\infty} 
\int_0^\infty ds' \tan e{\cal B}s' \; e^{\Phi(p',s')} \Bigg] \,.
\end{eqnarray}
The second term vanishes since $e^{\Phi(p,s)}$ vanishes at both limits
due to the term $-\epsilon|s|$ in it. The other exponential survives
only at the limit $s'=0$, and gives
\begin{eqnarray}
C_2 &=& \int \frac{d^4p}{(2\pi)^4} \eta_-(p) 
\int_{-\infty}^\infty ds \; \tan e{\cal B}s \; e^{\Phi(p,s)} 
\nonumber\\* 
&=& 0 \,,
\end{eqnarray}
where the last step follows on performing the integration over $p$,
since $\Phi(p,s)$ is an even function of $p$ and $\eta_-(p)$ is odd.

Let us now look at the other contribution, $C_1$. Separating out the
terms involving parallel components from those involving transverse
components, we write
\begin{eqnarray}
C_1 &=& \int \frac{d^4p}{(2\pi)^4} \eta_-(p)
\int_{-\infty}^\infty ds \int_0^\infty ds' \; 
\exp \big[ \Phi(p,s) + \Phi(p',s') \big] \nonumber\\* 
&\times & \Bigg[ (p_\parallel^{\prime2} - p_\parallel^2 ) 
\Big(\tan e{\cal B}s + \tan e{\cal B}s' - \; {\tan
e{\cal B}s \; \tan e{\cal B}s' \over \tan e{\cal B}(s+s')} \Big) 
\nonumber \\*
&& + p_\perp^2 \; \tan e{\cal B}s' \, \sec^2 e{\cal B}s -
p_\perp^{\prime2} 
\; \tan e{\cal B}s \, \sec^2 e{\cal B}s' \Bigg] \,.
\label{C1}
\end{eqnarray}
From the definition of $\Phi$, it follows that, apart from the small
convergence factors, 
\begin{eqnarray}
\Phi(p,s) + \Phi(p',s') &=& {i\over 2} (s+s') \left(
p_\parallel^{\prime2} + 
p_\parallel^2 - 2m^2 \right) - {i\over 2} (s-s') \left(
p_\parallel^{\prime2} - p_\parallel^2 \right) \nonumber\\*
&& - {i \over e{\cal B}} \left(
\tan e{\cal B}s'\; p_\perp^{\prime2} + \tan e{\cal B}s\; p_\perp^2
\right) \nonumber\\* 
&=& {i \over e{\cal B}} \Big[ \left( p_\parallel^{\prime2} +
p_\parallel^2 - 2m^2 \right) \xi - \left(
p_\parallel^{\prime2} - p_\parallel^2 \right) \zeta - 
p_\perp^{\prime2} \tan (\xi-\zeta) - p_\perp^2 \tan (\xi+\zeta) \Big] \,,
\nonumber\\*
\end{eqnarray}
where we have defined two new parameters
\begin{eqnarray}
\xi &=& \frac12 e{\cal B}(s+s') \,, \nonumber\\*
\zeta &=& \frac12 e{\cal B}(s-s') \,.
\label{xizeta}
\end{eqnarray}
Thus,
\begin{eqnarray}
{ie{\cal B}} \; {d\over d\zeta} e^{\Phi(p,s) + \Phi(p',s')} = 
e^{\Phi(p,s) + \Phi(p',s')} \left(
p_\parallel^{\prime2} - p_\parallel^2 - p_\perp^{\prime2} \sec^2
(\xi-\zeta) + 
p_\perp^2 \sec^2 (\xi+\zeta) \right) \,.
\label{C1par}
\end{eqnarray}
Using this, we can rewrite Eq.\ (\ref{C1}) as
\begin{eqnarray}
C_1 &=& \int \frac{d^4p}{(2\pi)^4} \eta_-(p)
\int_{-\infty}^\infty ds \int_0^\infty ds' \nonumber\\*
&\times& \Bigg[ \Big(\tan e{\cal B}s + \tan e{\cal B}s' - \; {\tan
e{\cal B}s \; \tan e{\cal B}s' \over \tan e{\cal B}(s+s')} \Big) 
ie{\cal B}\;{d\over d\zeta} e^{\Phi(p,s) + \Phi(p',s')} \nonumber\\* 
&& + e^{\Phi(p,s) + \Phi(p',s')} \Bigg\{
p_\perp^{\prime2} \tan e{\cal B}s' \sec^2 e{\cal B}s' 
\Big( 1 - \; {\tan
e{\cal B}s \over \tan e{\cal B}(s+s')} \Big) \nonumber\\*
&& \qquad \qquad \qquad - p_\perp^2 \; \tan e{\cal B}s \, \sec^2
e{\cal B}s 
\Big( 1 - \; {\tan e{\cal B}s' \over \tan e{\cal B}(s+s')} \Big) 
 \Bigg\} \Bigg] \,. 
\label{C1new}
\end{eqnarray}

We are now left with only the transverse components everywhere except
in the exponents. To write them in a useful form, we turn to Eq.\
(\ref{single_derivative}) and take another derivative with respect to
$p^{\alpha_\perp}$. From the fact that this derivative should also
vanish on $p$ integration, we find
\begin{eqnarray}
p_\perp^\alpha p_\perp^\beta \stackrel\circ= {1\over \tan e{\cal
B}s + \tan e{\cal B}s'} \Bigg[ -\,{ie{\cal B} \over 2}
g_\perp^{\alpha\beta} +  {\tan^2 e{\cal B}s' \over \tan e{\cal
B}s + \tan e{\cal B}s'} \;
k_\perp^\alpha k_\perp^\beta \Bigg] \,.
\end{eqnarray}
In particular, then,
\begin{eqnarray}
p_\perp^2 \stackrel\circ= {1\over \tan e{\cal
B}s + \tan e{\cal B}s'} \Bigg[ -ie{\cal B} +
{\tan^2 e{\cal B}s' \over \tan e{\cal B}s + \tan e{\cal B}s'} \;
k_\perp^2 \Bigg] \,.
\label{psq}
\end{eqnarray}
It then simply follows that
\begin{eqnarray}
p_\perp^{\prime2} \stackrel\circ= {1\over \tan e{\cal
B}s + \tan e{\cal B}s'} \Bigg[ -ie{\cal B} +
{\tan^2 e{\cal B}s \over \tan e{\cal B}s + \tan e{\cal B}s'} \;
k_\perp^2 \Bigg] \,.
\label{p'sq}
\end{eqnarray}
We now put these into Eq.\ (\ref{C1new}). After some straight forward
but cumbersome algebra, it is found that the terms involving $k$
cancel out, and we are left with
\begin{eqnarray}
C_1 &=& ie{\cal B} \int \frac{d^4p}{(2\pi)^4} \eta_-(p)
\int_{-\infty}^\infty ds \int_0^\infty ds' \nonumber\\*
&\times& \Bigg[ \Big(\tan e{\cal B}s + \tan e{\cal B}s' - \; {\tan
e{\cal B}s \; \tan e{\cal B}s' \over \tan e{\cal B}(s+s')} \Big) 
{d\over d\zeta} e^{\Phi(p,s) + \Phi(p',s')} \nonumber\\* 
&& - {e^{\Phi(p,s) + \Phi(p',s')} \over \tan e{\cal B}s + \tan e{\cal
B}s'} \Bigg\{ \tan e{\cal B}s' \sec^2 e{\cal B}s' 
\Big( 1 - \; {\tan
e{\cal B}s \over \tan e{\cal B}(s+s')} \Big) \nonumber\\*
&& \qquad \qquad \qquad \qquad - \tan e{\cal B}s \, \sec^2
e{\cal B}s 
\Big( 1 - \; {\tan e{\cal B}s' \over \tan e{\cal B}(s+s')} \Big) 
 \Bigg\} \Bigg] \,.
\end{eqnarray}
It is straight forward to show that this can be written in the
following form:
\begin{eqnarray}
C_1 &=& ie{\cal B} \int \frac{d^4p}{(2\pi)^4} \eta_-(p)
\int_{-\infty}^\infty ds \int_0^\infty ds' \; 
{d\over d\zeta} {\cal F}(\xi,\zeta) \,,
\end{eqnarray}
where
\begin{eqnarray}
{\cal F}(\xi,\zeta) =
\Big(\tan e{\cal B}s + \tan e{\cal B}s' - \; {\tan
e{\cal B}s \; \tan e{\cal B}s' \over \tan e{\cal B}(s+s')} \Big) 
e^{\Phi(p,s) + \Phi(p',s')} \,,
\end{eqnarray}
with $s$ and $s'$ related to $\xi$ and $\zeta$ through Eq.\
(\ref{xizeta}). 

We can now change the integration variables to $\xi$ and $\zeta$.
This gives
\begin{eqnarray}
C_1 &=& {2i\over e{\cal B}} \int \frac{d^4p}{(2\pi)^4} \eta_-(p)
\int_{-\infty}^\infty d\xi \int_{-\infty}^\infty d\zeta \;
\Theta(\xi-\zeta) \; {d\over d\zeta} 
{\cal F}(\xi,\zeta) \nonumber\\*
&=& {2i\over e{\cal B}} \int \frac{d^4p}{(2\pi)^4} \eta_-(p)
\int_{-\infty}^\infty d\xi \int_{-\infty}^\infty d\zeta \;
\Big[ {d\over d\zeta} \Big\{ \Theta(\xi-\zeta) \; 
{\cal F}(\xi,\zeta) \Big\} - \delta(\xi-\zeta) \; 
{\cal F}(\xi,\zeta) \Big] \nonumber\\*
&=& -\, {2i\over e{\cal B}} \int \frac{d^4p}{(2\pi)^4} \eta_-(p)
\int_{-\infty}^\infty d\xi \;
{\cal F}(\xi,\xi) \,,
\end{eqnarray}
since the other term vanishes at the limits. In this integrand,
$\zeta=\xi$, which means $s'=0$. Looking back at the definition of
$\cal F$, we find
\begin{eqnarray}
{\cal F}(\xi,\xi) = \exp \left\{\Phi(p,{2\xi\over e{\cal B}})\right\}
\tan 2\xi \,. 
\end{eqnarray}
This is an even function of $p$, whereas $\eta_-(p)$ is odd. Thus, the
expression vanishes on integrating over $p$.


\end{document}